\documentclass[twocolumn,prb,amsmath,showpacs,amssymb,citeautoscript,floatfix]{revtex4}
\usepackage{graphicx}
\def\etal{{ \it et al. }}
\def\prb{{Phys. Rev. B }}

\usepackage{dcolumn}
\usepackage{bm}
\usepackage{amsmath}
\begin{document}

\title{Magnetism of mixed quaternary Heusler alloys: 
(Ni,T)$_{2}$MnSn (T=Cu,Pd) as a case study}

\date{\today}

\author{S.K. Bose}
\affiliation{Department of Physics, Brock University,
St. Catharines, Ontario, Canada, L2S 3A1}

\author{J. Kudrnovsk\'y}
\affiliation{Institute of Physics, Academy of Sciences of the
Czech Republic, CZ-182 21 Praha 8, Czech Republic}

\author{V. Drchal} 
\affiliation{Institute of Physics, Academy of Sciences of the
Czech Republic, CZ-182 21 Praha 8, Czech Republic}

\author{I. Turek}
\affiliation{Charles University, Faculty of Mathematics and 
Physics, Department of Condensed Matter Physics, Ke Karlovu 5, 
CZ-12116 Prague 2, Czech Republic}

\date{\today}

\begin{abstract}
The electronic properties, exchange interactions, finite-temperature 
magnetism, and transport properties of random quaternary Heusler
Ni$_{2}$MnSn alloys doped with Cu- and Pd-atoms are studied
theoretically by means of {\it ab initio} calculations over
the entire range of dopant concentrations.
While the magnetic moments are only weakly dependent on the alloy
composition, the Curie temperatures exhibit strongly non-linear
behavior with respect to Cu-doping in contrast with an almost linear concentration
dependence in the case of Pd-doping.
The present parameter-free theory agrees qualitatively and also reasonably well
quantitatively with the available experimental results.
An analysis of exchange interactions is provided for a deeper understanding of the problem.
The dopant atoms  perturb electronic structure close to
the Fermi energy only weakly and the residual resistivity thus obeys a simple
Nordheim rule.
The dominating contribution to the temperature-dependent resistivity
is due to thermodynamical fluctuations originating from the spin-disorder, which, according to our
calculations, can be described successfully via the disordered local moments model. 
Results based on this
model agree fairly well with the measured values of  
spin-disorder induced resistivity.

\end{abstract}


\maketitle

\section{Introduction}
Heusler alloys were first studied by the German chemist Friedrich Heusler
 in 1903, starting with the ordered alloy  Cu$_2$MnSn. 
Because of their interesting physical properties  
they have been studied intensively in the past as well as more recently\cite{review}.
The most widely studied Heusler alloys are those with the formula
Ni$_2$MnZ (Z=Sn,Ga,In) and Co$_2$XY (X=Mn, Fe; Y=Al, Si).
The former group is of interest because of potential technological
applications based on their magnetic shape memory effect \cite{sme}, 
the magnetocaloric effect \cite{mce}, and the  recently observed giant 
(negative) magnetocaloric effect \cite{gmce}.
The latter group holds the promise of application 
 in spintronic devices, thanks to their halfmetallicity at room
temperature and above, lattice constant matching with the III-V semiconductors,
 and large bandgaps.  Large tunneling magnetoresistance was measured recently in Co$_2$MnSi/Al-O/Co$_2$MnSi magnetic tunneling junctions \cite{spintr}.

Structurally, most  Heusler alloys crystallize in two different 
cubic phases, having either the L2$_{\rm 1}$ (X$_2$YZ) or the 
C$_{\rm 1b}$ (XYZ) symmetry.
They can be best visualized as being composed of four interpenetrating 
fcc-sublattices, shifted along the body diagonal in the order X-Y-X-Z 
or X-Y-E-Z, where E denotes the empty, i.e.unoccupied, sublattice 
\cite{review}.
An important feature of Heusler alloys is the presence of chemical or
substitutional disorder.
It is often the non-stoichiometric composition with respect to
the ideal systems  such as Ni$_{2}$MnSn or  Ni$_{2}$MnSb which
interpolates between the L2$_{1}$ Heusler and the C$_{\rm 1b}$
semi-Heusler alloys \cite{ha-sha}. Examples are the magnetic shape memory alloys of the
type Ni$_{2}$Mn$_{1+x}$Sn$_{1-x}$ (Mn-nonstoichiometry), or the 
Ni$_{2-x}$MnSb (Ni-nonstoichiometry) alloys.

In addition to the chemical disorder due to nonstoichiometry,
a native chemical disorder exists even in 'ideal' ordered alloys
X$_2$YZ and XYZ.
In particular, halfmetallic Heusler alloys are very susceptible to such
native disorder, a typical example being the Co$_{2}$MnSi alloy
which exists in the B2-like structure due to Mn-Si disorder.

Finally, there are complex quaternary alloys like  the
semi-Heusler (Ni,Cu)MnSb alloys \cite{qsha1,qsha2} and closely 
related Heusler alloys like (Ni,Cu)$_{2}$MnSn \cite{qha1} 
or (Ni,Pd)$_{2}$MnSn \cite{qha2}, with disorder on the X-sublattice.
The magnetic properties of these alloys are simplified by the fact
that the Mn-sublattice is effectively nonrandom (although there can be
 some magnetic disorder in a certain concentration range
\cite{qsha1}). In addition, Mn-atoms carry essentially all of the system magnetic 
moment. However, despite these simplifying features the magnetic, thermodynamic, and transport
properties of these alloys are not easy to understand. Several factors: disorder on 
sublattices neighboring the unperturbed Mn-sublattice,  varying 
carrier concentration, the presence of atoms with
different degrees of $d$-electron localization and thus different levels of hybridization 
with Mn-atoms, all contribute to the complexity of the problem.
A reasonably successful interpretation of the results of
 experiments in Refs.[~\onlinecite{qha1,qha2}] 
was provided by Stearns \cite{stearns1,stearns2} using a qualitative model.
According to this, magnetic behavior of Heusler alloys is
controlled by three types of magnetic interactions, namely: 
(i) interaction between extended $s$-like electrons and localized $d$-electrons
mediated by the $sd$-hybridization, (ii) interaction between
localized and delocalized (itinerant) $d$-electrons, and (iii) the
superexchange mediated by $sp$-element on the Z-sublattice (Sn in
the present case).
The dominating interaction is that between the localized and itinerant 
$d$-electrons and as such the
value of the Curie temperature $T_c$ of Heusler alloys is essentially controlled by 
the amount of itinerant $d$-electrons.  
This, in turn, is related to the 
localization of the corresponding $d$-orbitals  of the X-atoms.
All of the above mentioned interactions are naturally present in the 
first-principles description of magnetism in the framework of the 
spin density functional theory (DFT). In general, it is very difficult to separate out
the individual contributions from the calculated total interactions.

In this work we employ  state-of-the-art
electronic structure calculations to understand the properties of 
disordered quaternary Heusler alloys (Ni,Cu)$_{2}$MnSn and 
(Ni,Pd)$_{2}$MnSn.  
The exchange interactions are computed
from the \textit{ab initio} electronic structure, and
then used to estimate the $T_c$ using various approximations.
Such a program has been carried out  in recent years for a number of 
conventional (ordered) Heusler alloys.
Notable are the pioneering work of K\"ubler \cite{kuebler1} 
and extensive studies of electronic and magnetic properties of
Heusler alloys, including  estimates of the Curie temperature, 
by K\"ubler \cite{kuebler2}, Galanakis \cite{gala}, 
Picozzi \cite{picozzi}, and Sasioglu and Sandratskii \cite{sandr}. 
The study of disordered Heusler alloys is, however, much more 
involved from a theoretical standpoint  because of the presence of disorder 
violating the translational symmetry.
Although in some cases, such as the Heusler alloys with $sp$-disorder on
the Z-sublattice, it is possible to use conventional band structure methods
(the supercell approach \cite{ha-sc}), more sophisticated methods, typically
using the coherent potential approximation (CPA) implemented within 
DFT formalism (DF-CPA), need to be employed \cite{ha-cpa1,ha-cpa2} to treat disorder in
general.
This is particularly true for systems with  general compositions
such as the magnetic shape memory alloys or the 
 Heusler alloys in  narrow concentration ranges where abrupt changes in physical properties are
known to occur (e.g. around the austenite-martensite transition). So far, such studies
have been typically limited to the electronic structure 
and  simple magnetic properties, e.g. magnetic moments.
 The first attempts to study  thermodynamical properties
(including $T_c$) of random alloys have appeared \cite{ha-sha,qsha2} only recently.
 Particularly worth mentioning is the extensive study of the semi-Heusler alloys
(Cu,Ni)MnSb \cite{qsha1} comprising the electronic structure, magnetic, 
thermodynamical, and transport properties of random alloys in the 
framework of a unified DFT description \cite{qsha1}.

The aim of the present work is to study electronic, magnetic, thermal,
and transport properties of two related disordered Heusler quaternary 
systems  based on the reference alloy Ni$_{2}$MnSn, namely the (Ni,Cu)$_{2}$MnSn 
and the (Ni,Pd)$_{2}$MnSn systems. 
Both systems have magnetic moments around 4 $\mu_{\rm B}$ over the
whole concentration range but dramatically different concentration
dependence of their $T_c$: strongly nonlinear in the former
case and almost linear in the latter.
The  $T_c$ of end-point alloys, namely X$_{2}$MnSn,
with X=Cu, Ni, and Pd, have decreasing values in this order. Thus they seem to
 obey the criterion advanced by Stearns\cite{stearns1} for the dependence of 
 $T_c$ on the localization of $d$-electrons
of the element X. Localization of the $d$-electrons increases left to right across a transition metal row,
and decreases across a column as we go from 3- to 4- and then to 5-$d$. Thus Cu $d$-electrons are more localized than
Ni $d$-electrons, which are more localized than Pd $d$-electrons, explaining the   
differences in $T_c$ of the three systems in the order mentioned above.

Transport properties of Heusler alloys have not been studied extensively. 
Some measurements of the temperature dependence of resistivity have been 
carried out\cite{resT1,resT2}. 
To our knowledge, no theoretical studies of  transport in Heusler alloys have
 appeared so far. 
We attempt to fill this gap, using a simplified approach.

\section{Formalism}
\label{Form}

The electronic structure calculations were performed
using the tight-binding linear muffin-tin orbital (TB-LMTO)
scheme \cite{lmto} in the framework of the local density
approximation (LDA).
The effect of substitutional disorder on the X-sublattice (either
Ni-Cu or Ni-Pd) is described by 
CPA formulated in the framework  of the TB-LMTO Green's function  method \cite{book}.
The same atomic sphere  radius was used for  all the constituent atoms, 
and lattice constants were taken from 
experiments \cite{qha1,qha2}. The calculations employed  an $s,p,d,f$-basis.
For the parameterization of the local density functional
the Vosko-Wilk-Nusair exchange-correlation potential \cite{VWN}
was used.

The thermodynamical properties of the system are assumed to be given by
a classical Heisenberg Hamiltonian,
\begin{equation}\label{e1}
H_\mathrm{eff} = - \sum_{i,j} \, J_{ij} \,
{\bf e}_{i} \cdot {\bf e}_{j} \ ,
\end{equation}
where $i,j$ are site indices, ${\bf e}_{i}$ is the unit vector
pointing along the direction of the local magnetic moment at
site $i$, and $J_{ij}$ is the exchange integral between
sites $i$ and $j$.
The exchange integrals, by construction, contain the atom magnetic
moments, their positive (negative) values being indicative of
ferromagnetic (antiferromagnetic) coupling.

We evaluate exchange integrals in Eq.~(\ref{e1}) using
 a two-step model \cite{lie,eirev}, where the band energy is equated to the Heisenberg form and
then expressed via multiple scattering formalism based on the TB-LMTO-ASA Green's function
in terms of the moments directed along ${\bf e}_{i}$ and ${\bf e}_{j}$.
The reference state for this calculation is chosen to be the
disordered local moment (DLM) state \cite{dlm}.  
Such a choice was recently suggested for the study of semi-Heusler alloys \cite{qsha2}, as 
it has some advantages over the conventional choice of the ferromagnetic 
reference state \cite{lie}.
The DLM state is closer to the state at which the magnetic 
transition occurs, compared with  the state with a global 
magnetization.  There is no preferred magnetic configuration 
assumed, and there are no induced moments for the DLM state 
(see recent discussion of the problem of induced moments in 
the Heisenberg model in Ref.~\onlinecite{induced}). The DLM reference 
 state was  successfully used
recently in the study of magnetic overlayers on non-magnetic
substrates \cite{over}. Ideally one would hope the resulting magnetic properties to be
robust, i.e. (almost) independent of the assumed reference state.
In practice, some dependence on the reference state is unavoidable, and 
 in some cases the DLM reference state has been found to provide
better estimates of the Curie temperatures.

We determine the Curie temperature
corresponding to the effective Heisenberg model in Eq. (\ref{e1}) by
making use of the random-phase  approximation (RPA).  For comparison,  we
also include results obtained in the framework of the the mean-field 
 approximation (MFA).
The RPA-Curie temperatures are known to be close to those obtained from
Monte-Carlo simulations \cite{eirev}.

We have used interactions up to $\sim$ 4 lattice
constants which are found to be sufficient to achieve reasonably converged results.
While it is not a problem to include even more distant interactions in the
evaluation of the RPA-Curie temperature, problems arise
in connection with integration of the inverse lattice Fourier
transform of real-space exchange integrals over the Brillouin zone
which becomes a  rapidly oscillating function for distant shells.
As a result, the RPA Curie temperature oscillates as a function
of the shell number included in the above mentioned lattice Fourier
transform.
It should be noted that such problems do not arise for the MFA or while dealing with 
semi-Heusler alloys \cite{qsha1,qsha2},  where the exchange integrals are
strongly damped in real-space due to their halfmetallicity 
\cite{ha-sha}.
We have adopted the approach proposed in Ref.~\onlinecite{pajda} and 
used in the context of Heusler alloys in Ref.~\onlinecite{ha-sha}.
It relies on using a set of exponential damping parameters to compute the $T_c$, thereby reducing the 
oscillations and finally extrapolating the results to the zero 
damping case.
This approach was successfully applied to the evaluation of the
stiffness constant of ferromagnets in  real-space, where the problems associated
with using only a limited number of shells are even more severe \cite{pajda}. 
The above-mentioned oscillations are even stronger
for  the ferromagnetic reference state.

Without an external magnetic field, there are essentially three
different contributions to the resistivity of magnetic alloys:
(i) phonon scattering, (ii) magnetic scattering due to 
thermodynamical fluctuations, which are largest close to
the Curie temperature \cite{magsc}, and (iii) the residual 
resistivity due to the presence of chemical disorder on the 
Ni-Cu and Ni-Pd sublattices.

In this study we limit ourselves to the residual resistivity, and also
present  a simplified treatment of the resistivity due to magnetic (spin-disorder) scattering. 
The residual resistivity is determined by the linear-response 
theory as formulated in the framework of the TB-LMTO-CPA approach using the
Kubo-Greenwood formula \cite{kglmto}, i.e., on the same formal
footing as used for the determination of the exchange integrals.
This approach, formulated for the multi-sublattice case, 
allows us to include both the substitutional disorder on the
Ni-sublattice as well as magnetic disorder on the
Mn-sublattice on an equal footing \cite{trit}.
The disorder-induced vertex corrections are  included in
the formalism \cite{carva}.
The spin disorder is described here in the framework of the
DLM, and treated formally as  'substitutional' disorder via  CPA.
We refer the reader to Ref.~\onlinecite{book} for details on the
implementation of the DLM in the framework of the TB-LMTO 
formalism.

\section{Results and discussion}

In this section we present results for the electronic, magnetic,
 and transport properties of (Ni,T)$_{2}$MnSn
(T=Cu, Pd) alloys over a broad range of concentrations and
compare our results with available experimental data.

\subsection{Density of states}
\label{DOS}

\begin{figure*}
\center \includegraphics[width=18cm]{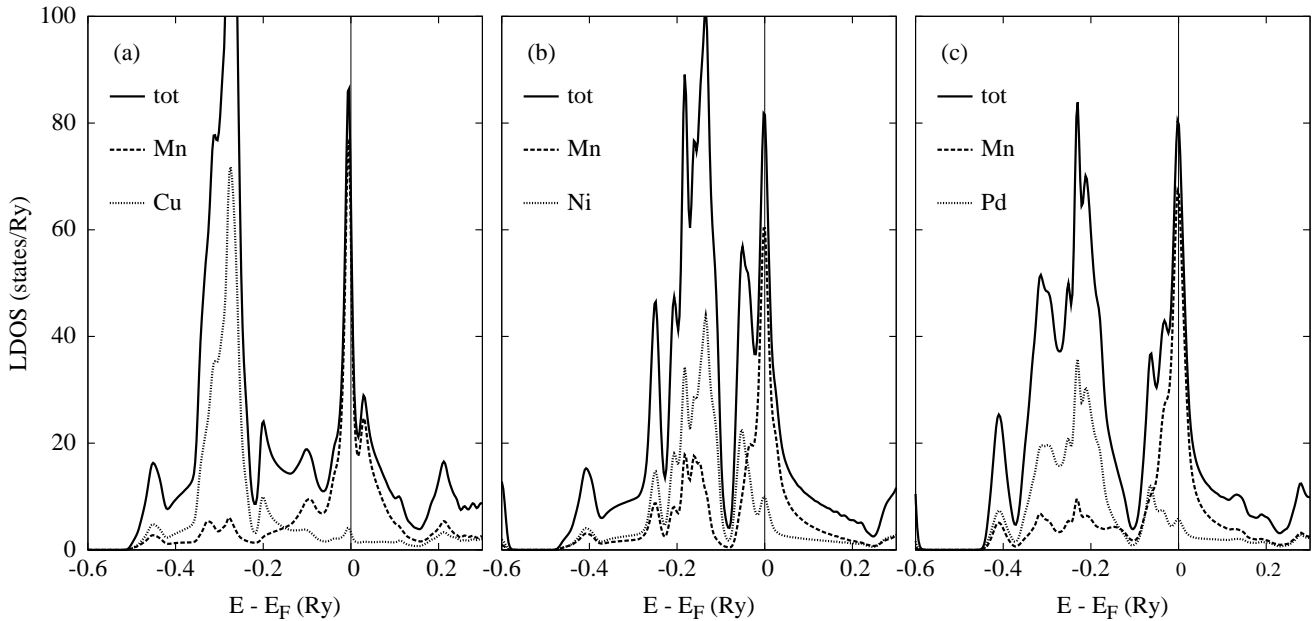}
\caption {Total and component resolved densities of states, 
per formula unit and spin,
 for nonmagnetic Heusler alloys: 
(a) ordered Cu$_{2}$MnSn, 
(b) ordered Ni$_{2}$MnSn, and (c) ordered Pd$_{2}$MnSn.
}
\label{f1}
\end{figure*}

 Fig.~\ref{f1} shows the calculated local densities of states 
(LDOSs) of nonmagnetic X$_{2}$MnSn (X=Cu, Ni, Pd) alloys.
Several conclusions follow immediately: (i) all DOSs have a pronounced
peak at the Fermi energy ($E_{\rm F}$),  indicating  instability 
against formation of the ferromagnetic state (Stoner criterion).
(ii) The dominant contribution to the peaks is from the Mn-LDOS, 
which suggests that the magnetism is primarily due to the
Mn-sublattice. 
(iii) The height of Mn-LDOS($E_{\rm F}$) (which is $\sim$ total DOS($E_{\rm F}$)) is approximately the same
in all cases,  indicating similar magnetic moments for all these 
compounds (see Fig.~\ref{f3} below).
(iv)The increasing bandwidth of the $d$-states dominating local 
X-LDOS in the order Cu-Ni-Pd is clearly seen. This is indicative of decreasing localization
(increasing itinerancy)  of the $d$-electrons, resulting in decreasing $T_c$ in the same order,
according to the model proposed by Stearns \cite{stearns1}.
(v) The Cu- and Ni-bands are well separated in energy resulting
in a strong diagonal disorder, while their bandwidths are
approximately the same (weak off-diagonal disorder). On the contrary, Ni- and Pd-bands differ
mainly in their widths (off-diagonal disorder). In all
cases, however, X-bands are well separated energetically from the
Fermi energy so that  disorder on the X-sublattice influences
states at $E_{\rm F}$ only weakly. Hence, the residual resistivity 
(due to chemical disorder) should be in the weak-scattering regime.

\begin{figure*}
\center \includegraphics[width=18cm]{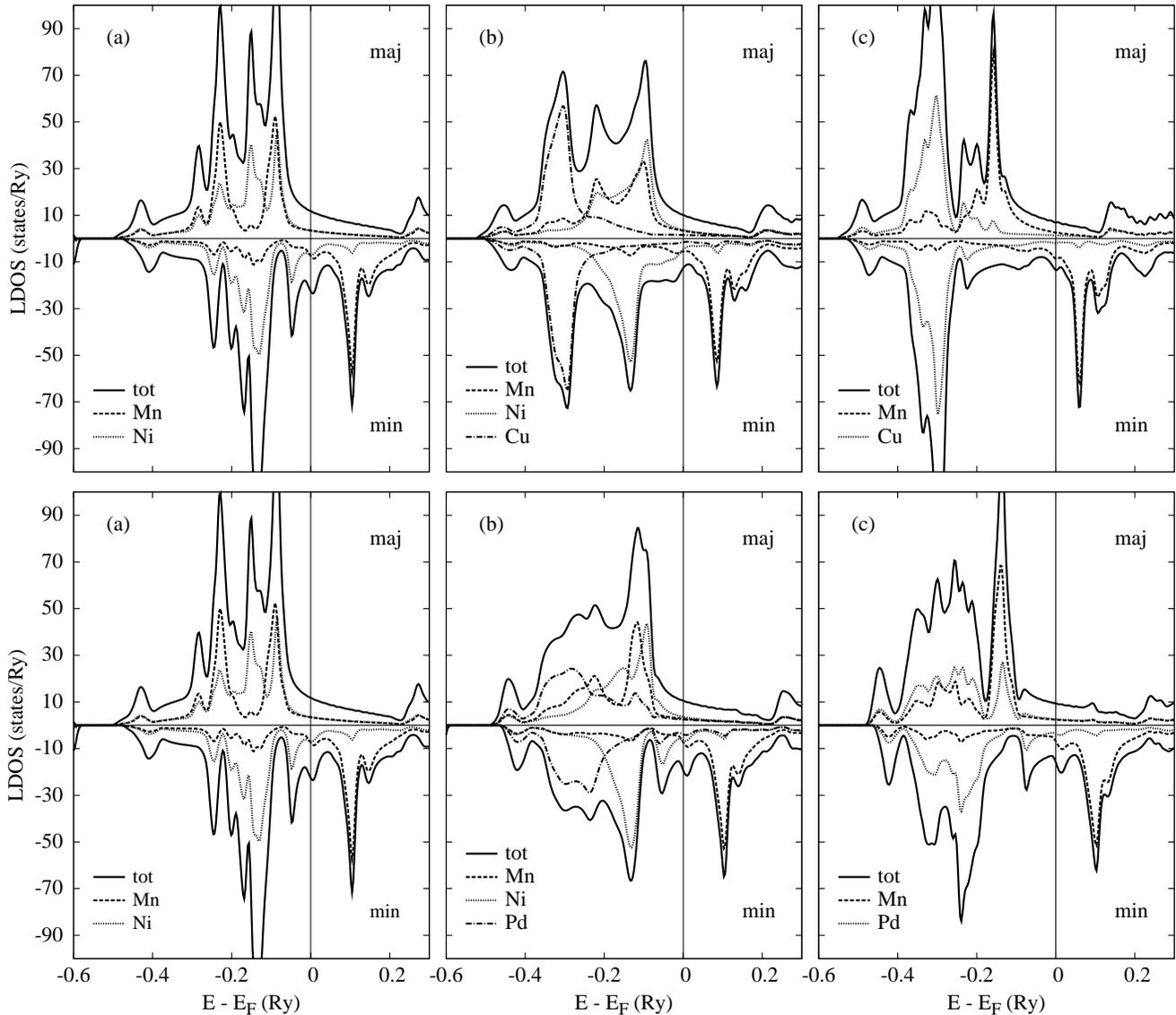}
\caption {Total and component resolved densities of states,
per formula unit and spin,
 for
ferromagnetic Heusler alloys. Top panel: (a) ordered Ni$_{2}$MnSn,
(b) disordered (Ni$_{50}$,Cu$_{50}$)$_{2}$MnSn, and (c) ordered 
Cu$_{2}$MnSn.
Bottom panel: (a) ordered Ni$_{2}$MnSn, (b) disordered 
(Ni$_{50}$,Pd$_{50}$)$_{2}$MnSn, and (c) ordered Pd$_{2}$MnSn.
Majority (minority) densities of states are shown in upper (lower)
parts of each figure.
}
\label{f2}
\end{figure*}

The spin-polarized LDOS for Ni$_{2}$MnSn, (Ni$_{50}$,Cu$_{50}$)$_2$MnSn, 
and Cu$_{2}$MnSn are shown in Fig.~\ref{f2} (top panel).
The following conclusions can be drawn: (i) Spin-polarized Cu-LDOSs 
in Cu$_{2}$MnSn are almost identical, indicating practically no 
polarization, while some small polarization is seen for corresponding
Ni-LDOS in Ni$_{2}$MnSn. Similar conclusions are also valid for
Cu- and Ni-LDOSs for the equiconcentration case (Fig.~\ref{f2}b).
(ii) The total LDOS is smoothed by strong level disorder 
in the  equiconcentration alloy. (iii) Due to  large level splitting
(large local magnetic moment) the minority Mn-bands in all cases 
 hybridize very little with the X-bands. On the contrary, such hybridization,
compared to the nonmagnetic case, is strong
for the majority bands   even for Cu$_{2}$MnSn. (iv) The majority and minority states at 
the Fermi energy behave differently (corresponding DOSs have different curvature). This results in
different Fermi surface geometry for these bands (see e.g.  Ref.~\onlinecite{fermi}).
The results for Ni$_{2}$MnSn, (Ni$_{50}$,Pd$_{50}$)$_2$MnSn, and 
Pd$_{2}$MnSn are shown in Fig.~\ref{f2} (bottom panel).
There are some differences in this case: (i) The carrier concentration
is the same through the entire alloy system. As a result, the
position of the minority Mn-band with respect to the Fermi level stays fixed.
This is different from the (Ni,Cu)$_{2}$MnSn alloy system where the
 carrier concentration increases with Cu-content.  (ii) We observe
strong hybridization of Ni- and Pd-bands on the X-sublattice, which
differ mostly in their widths. (iii) One thus expects weaker site
off-diagonal disorder  and, consequently, also
lower residual  resistivity (see Section~\ref{RR} below).

\subsection{Magnetic moments}
\label{MM}

\begin{figure}
\center \includegraphics[width=8.5cm]{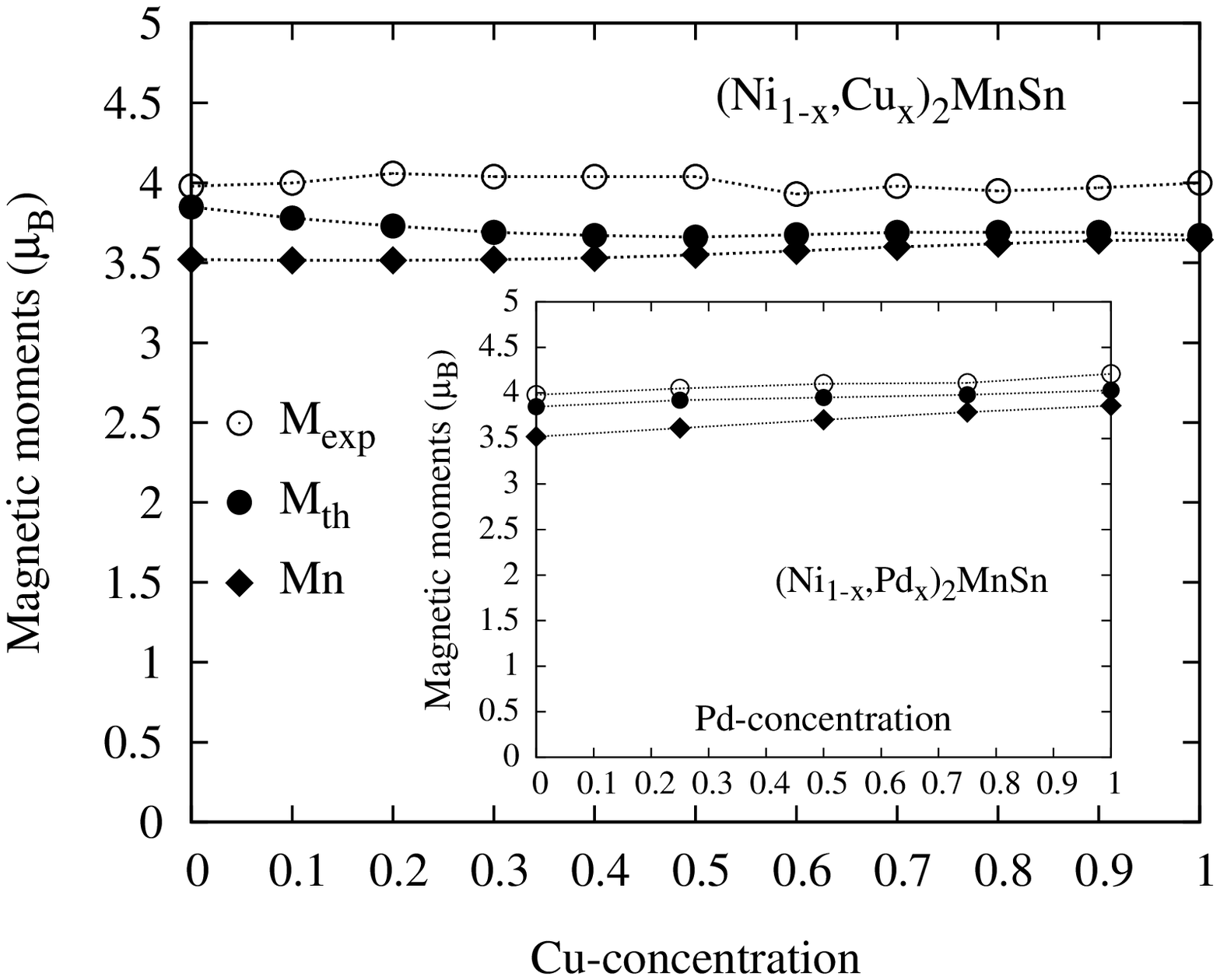}
\caption {Averaged magnetic moment (full circle) and local magnetic 
moments of Mn-atoms (diamonds) as a function of composition for 
(Ni,Cu)$_{2}$MnSn alloys. Related results for (Ni,Pd)$_{2}$MnSn
are shown as inset.  Experimental data \cite{qha1,qha2} are
denoted by empty circles.
}
\label{f3}
\end{figure}

Experimental and calculated average moments per formula unit (f.u.) are 
presented in Fig.~\ref{f3}, together with calculated local moments 
on Mn-sublattices. 
There is an overall good agreement between calculated and measured
average moments which are essentially concentration-independent
and have values around $4~\mu_{\rm B}$.  
For the stoichiometric alloys X$_2$MnSn (X = Ni, Cu, Pd),
very similar values were obtained in a previous study
by the TB-LMTO-ASA method.\cite{kurtulus2005}
Mn-sites carry almost all the moment per f.u. and  this is particularly true for
Cu-rich alloys.
Although the calculated magnetic moments agree with experimental 
values within a few percent, even better agreement can be obtained
(e.g. for Ni$_{2}$MnSn) when calculations are performed using
the generalized gradient approximation (GGA) \cite{sandr} instead of LDA.
The weak concentration dependence of the averaged magnetic moment can
be understood from Fig.~\ref{f2}. 
The large intra-atomic  exchange splitting of Mn atoms and the small
hybridization of minority Mn-d orbitals with the X-atom orbitals
lead to the full occupation of majority Mn-d orbitals, to
negligible spin polarization of X- and Sn-atom orbitals, and to
composition-independent occupation of minority Mn-d orbitals.
These features result in the nearly constant Mn-moment and the total
alloy magnetization.
The local moments on Ni-sites have values in the range 
(0.14, 0.19)~$\mu_{\rm B}$ for both systems and all concentrations.
The local Cu moments are very small (0.01$-$0.02~$\mu_{\rm B}$),
while local Pd-moments are almost concentration-independent, with  
a value $\sim$ 0.1~$\mu_{\rm B}$.
Finally, local moments on the Sn-sublattice are small and negative,
in the range (-0.02, -0.04)~$\mu_{\rm B}$.

\subsection{ Exchange interactions}
\label{EI}
\subsubsection{ Calculated results}
\begin{figure*}
\center \includegraphics[width=18cm]{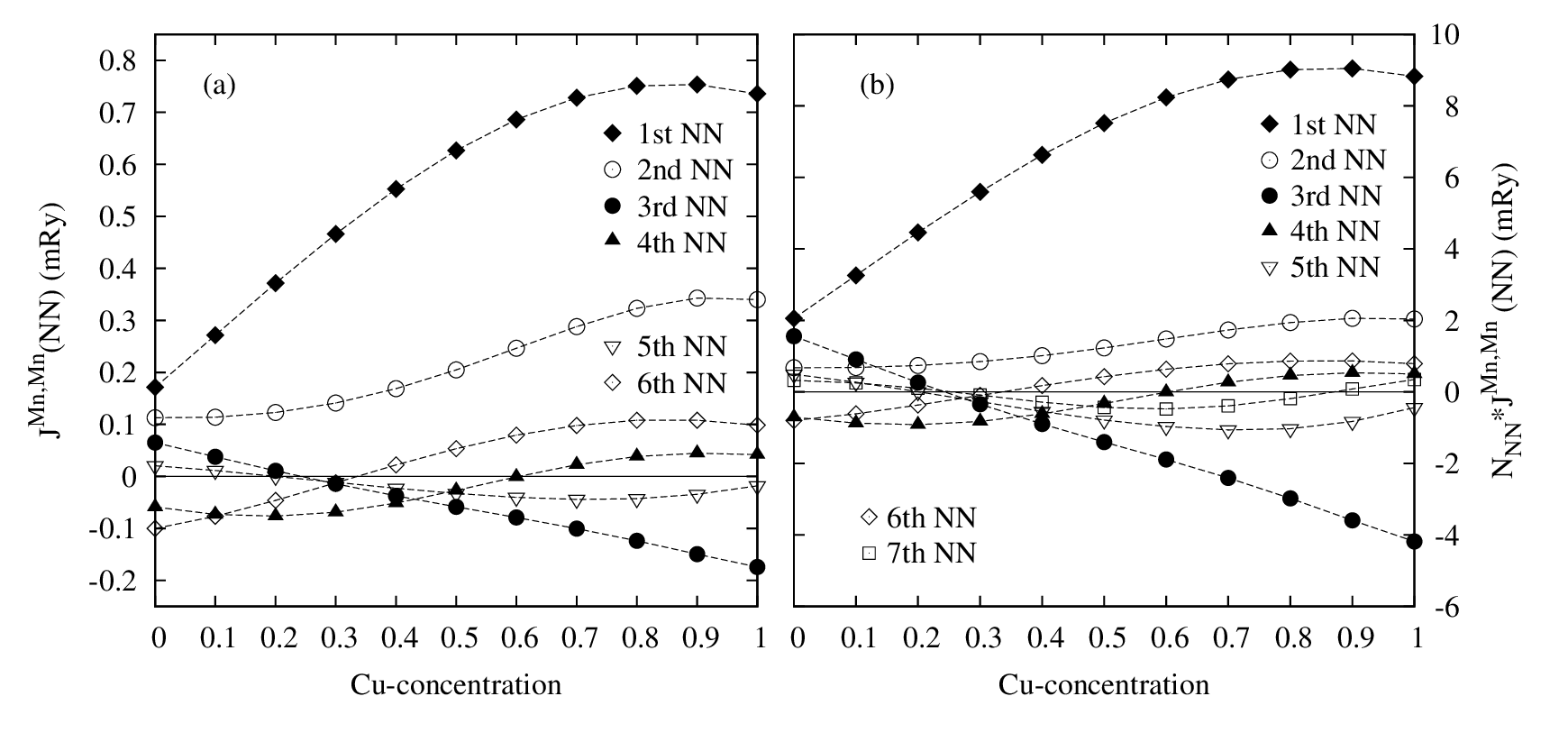}
\caption {Concentration dependence of exchange integrals
in (Ni$_{1-x}$,Cu$_{x}$)$_{2}$MnSn Heusler alloys as estimated from
the disordered local magnetic reference state: (a) the first six
leading (Mn,Mn)-exchange integrals, and (b) the same but with
exchange integrals multiplied by their shell degeneracies.
We have also added the 7th shell term with a large degeneracy.
}
\label{f4}
\end{figure*}

The exchange integrals are one of the important characteristics of
the magnetically polarized state.
We have determined them as a function of the alloy composition 
using the DLM reference state.
 It should be noted that for the DLM reference state the only 
non-zero exchange interactions are those between the Mn sites.
The Mn-moments are very rigid and their values in the ferromagnetic
and DLM states almost coincide with each other.
The leading exchange integrals in (Ni$_{1-x}$,Cu$_{x}$)$_{2}$MnSn 
quaternary Heusler alloys are plotted as a function of the 
Cu-concentration in Fig.~\ref{f4}a.
The Mn-sublattice is  only indirectly perturbed by the alloy
disorder on the X-sublattices, via the hybridization
of Ni- and Cu-atoms with the Mn-sublattice.
Such a hybridization is  weak for Cu-atoms, but strong for the
Ni-atoms (see Figs.~\ref{f1} and \ref{f2}).
In addition to Mn-X hybridization, which varies with the alloy
composition, there is also an increase of the total electron concentration
with the Cu-content due to different valency of Ni- and Cu-atoms.
The resulting concentration dependence is quite complex, with the
first three exchange integrals being dominant.
While the first two are ferromagnetic (FM) and increase with the Cu-content,
the third one decreases with Cu-concentration almost linearly and 
changes its sign from  FM to antiferromagnetic (AFM) at about 
20\% of Cu.
It should be noted, however, that  the 4th and in particular
the 6th neighbor interactions  are also non-negligible.
To better understand the contribution of the various shells to the 
magnetic properties we have multiplied the exchange integrals by their degeneracies
(N), i.e., by the number of the equivalent atoms in the corresponding
shells.
This is presented in Fig.~\ref{f4}b.
Such terms provide the actual contribution
of a given shell to the MFA expression of the Curie temperature 
\cite{lie,eirev}.
With the multiplication by the degeneracy factors, the contribution of the first shell remains the largest one,
and the contributions of the 2nd and 6th shells are suppressed compared to that of the 3rd shell. 
We have also shown the contribution of the 7th shell, which is non-negligible
 due to its large degeneracy, even though individual atom contribution is very small.

The above results confirm the assumptions of a theoretical model of 
Stearns \cite{stearns2}, which asserts that at least three independent magnetic 
interactions are needed in order to explain the magnetic properties
of Heusler alloys with late 3 $d$ and 4 $d$ elements on the
X-sublattices.
 This is a probable cause for the failure of a first-principles 
study \cite{kuebler1} in which the authors tried to estimate the 
Curie temperature from just two exchange integrals,
based on the energy differences between the FM state and two
different AFM states (AF~I and AF~II with antiparallel alignments 
of spins along the [100] and [111] directions).
A bad estimate of the Curie temperature for such a model for 
Cu$_{2}$MnSn can be attributed particularly to the large contribution of
the 3rd shell (see Fig.~\ref{f4}b), which was ignored.

\begin{figure*}
\center \includegraphics[width=18cm]{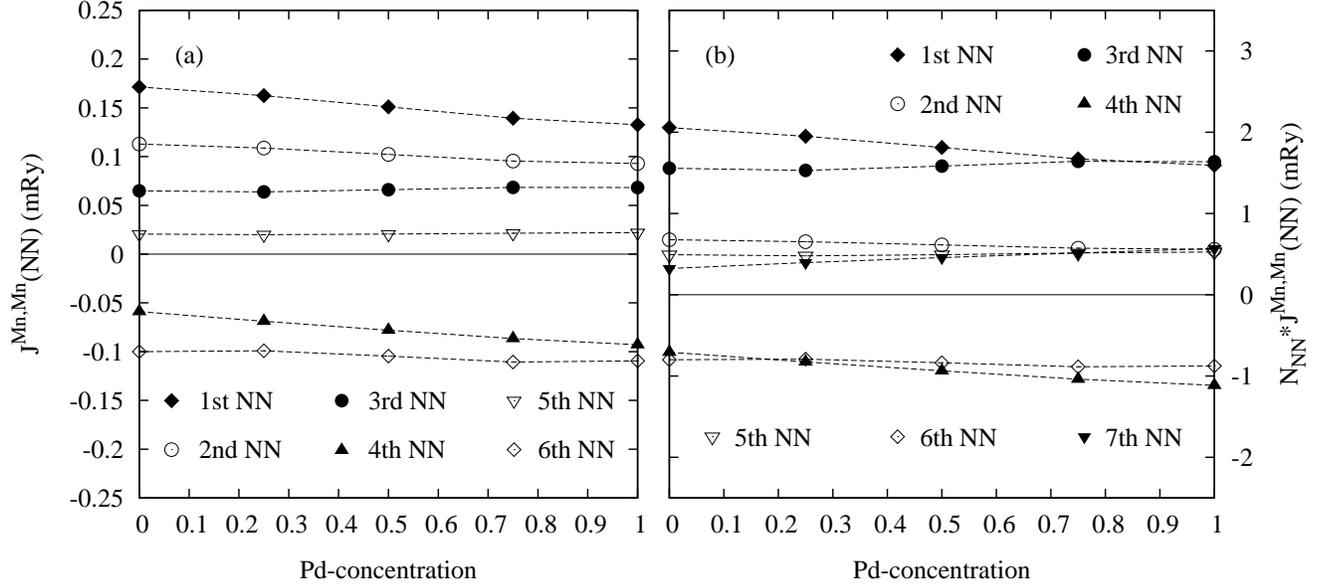}
\caption {The same as in Fig.~\ref{f4} but for
(Ni$_{1-x}$,Pd$_{x}$)$_{2}$MnSn Heusler alloys.
}
\label{f5}
\end{figure*}

In  Fig.~\ref{f5} we present the results of a similar study for 
(Ni$_{1-x}$,Pd$_{x}$)$_{2}$MnSn Heusler alloy.
There are some important differences with respect to the (Ni,Cu)MnSn alloy: 
the number of valence electrons 
remains unchanged upon alloying, Ni and Pd being neighbors
in the same column of the Periodic Table.
As already mentioned, the effect of disorder is smaller,
as it has predominantly off-diagonal character.
The leading interactions are FM, but some interactions, e.g. 
those from the 4th and 6th shells, are AFM (see Fig.~\ref{f5}a).
Upon multiplying the exchange integrals
by their shell degeneracy (Fig.~\ref{f5}b),
 the contributions from the 1st and 3rd shell are seen to be dominant,
while the contributions from the 2nd, 5th, and 7th shells (FM) and
those of 4th and 6th shells (AFM) tend to compensate each
other in the MFA sum.
The contributions of all shells are either almost concentration-independent 
or decrease slightly  with the Pd-content, resulting in 
 a linear decrease of the Curie temperature with
Pd-concentration (see below).

\begin{figure}
\center \includegraphics[width=8.5cm]{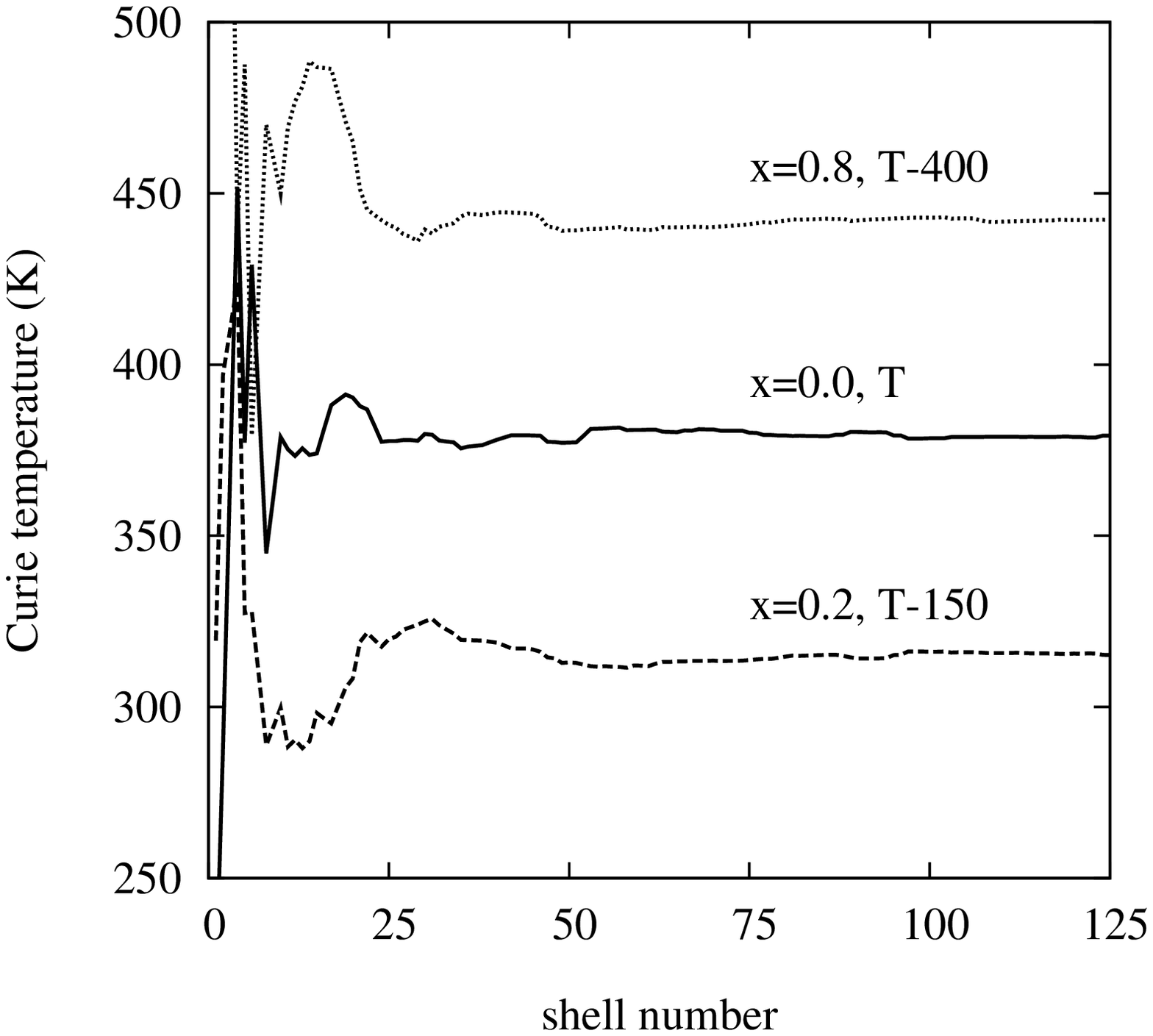}
\caption {The mean-field Curie temperature of 
(Ni$_{1-x}$,Cu$_{x}$)$_{2}$MnSn for $x$=0, 0.2, and 0.8 
illustrating the fulfillment of the saturation of results with respect to the
number of shell. Note that 
curves for $x$=0.2 and 0.8 were shifted to fit the
frame.
}
\label{f6}
\end{figure}

There is a natural question as to how many shells are needed for a
reliable estimate of the Curie temperature.
We have just seen that the smallness of exchange integral is not
a sufficient condition, as it can be offset by a large degeneracy of the shell.
We have insured
that the weighted sum of exchange integrals as a function of 
the shell number achieves its saturation value.
 For halfmetallic alloys with exponentially
damped exchange integrals \cite{ha-sha,qsha1,qsha2} this condition
is fulfilled for a  relatively small number of shells.
 In transition metal alloys with strong level disorder
(see e.g. Ref.\onlinecite{magall}) the situation is similar, because
the disorder influences the states at the Fermi energy, which are the relevant 
states dictating  
the spatial dependence of exchange integrals \cite{eirev}.
However,  for the Heusler alloys considered here,  the effect of disorder at the
Fermi energy is not as strong as in halfmetallic alloys. 
Hence a larger number of shells needs to be included.
We illustrate the situation in Fig.~\ref{f6}, where we plot the
mean-field Curie temperature for three typical compositions in 
(Ni$_{1-x}$,Cu$_{x}$)$_{2}$MnSn alloy.
We note that the mean-field value of the  Curie temperature is 
directly proportional to the sum of the exchange interactions \cite{lie,eirev}.
We observe that 
for the shell distance $d \le $ 4~$a$, where $a$ is 
the lattice constant, the results are reasonably saturated.
We have thus used all neighbors up to 4~$a$ and applied the
extrapolation procedure briefly described in the Introduction
\cite{pajda}.
This procedure is necessary, as the RPA $T_c$ values oscillate more strongly than the MFA
values due to the contribution from the neighborhood of the
zone center, where the contribution of the distant shells plays an important role.
The estimated error of this procedure is a few percent of the
absolute value of the Curie temperature.

\begin{figure}
\center \includegraphics[width=8.5cm]{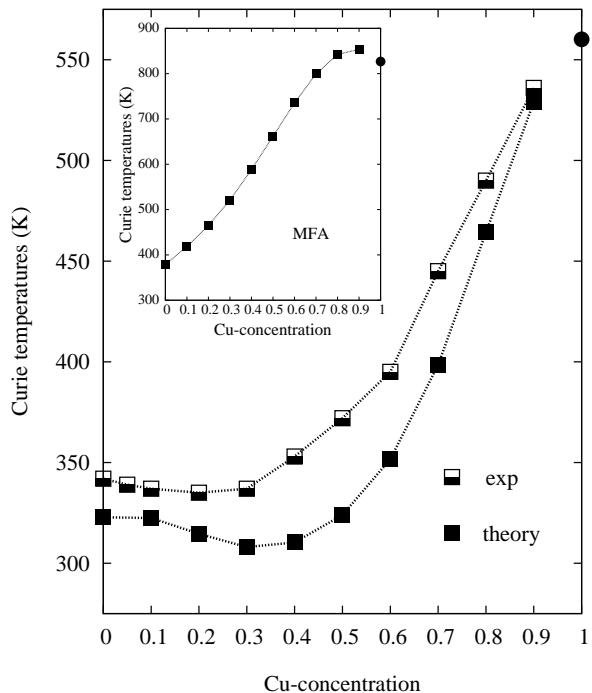}
\caption {The concentration dependence of Curie temperatures 
for (Ni$_{1-x}$,Cu$_{x}$)$_{2}$MnSn. The calculated values (RPA)
are compared with experimental data \cite{qha1}. The MFA
values of Curie temperatures are shown in the inset.
Filled circles denote the calculated MFA and RPA Curie temperatures for
 Cu$_{2}$MnSn.}
\label{f7}
\end{figure}

Calculated results  for (Ni,Cu)$_{2}$MnSn  are presented and compared
with experiment \cite{qha1} in Fig.~\ref{f7}.  

The system (Ni$_{1-x}$,Cu$_{x}$)$_{2}$MnSn exhibits different 
behaviors in two concentration regions: for $x \in (0.0,0.3)$  
the Curie temperature stays almost constant with a small minimum  
at $x \approx 0.3$, while for $x \ge 0.4$ it increases monotonically. 
Theoretical calculations in the framework of the RPA 
reproduce the experimental data reasonably well, both qualitatively and 
quantitatively.
In contrast, the MFA not only overestimates the Curie temperature,
but also exhibits a monotonic increase in the whole concentration
range, in contrast to the experiment.
According to Stearns \cite{stearns1}, the larger $T_c$ of
Cu$_{2}$MnSn  compared to that of Ni$_{2}$MnSn is due to stronger localization
of $d$-electrons, which enhances the Curie temperature.
The flat minimum is explained by the competition of this effect with
the weakening of this interaction for larger interatomic distances
(larger lattice constant) with increasing Cu-content \cite{qha1}.
It is obvious that this behavior is due to AFM-like
exchange interactions of the 3rd shell. For low Cu-concentration this interaction
somewhat reduces the RPA sum, and for higher Cu-content  its
effect is compensated by increasing FM-like interactions. 

The calculated Curie temperature for Cu$_{2}$MnSn
 is also shown in Fig.~\ref{f7}.
Experimental values are not discussed in Ref.~\onlinecite{qha1}.  
There is some controversy in the literature: values 530~K and 630~K 
can be found, although the lower one is sometimes questioned. 
The calculated RPA value 560~K lies in the range of the 
reported experimental values.

The results for Ni$_{2}$MnSn alloy can be compared with 
other theoretical work \cite{ni2mnsn}.
The experimental values of the Curie temperature range from 
328~K to 360~K.
The Curie temperature estimated in Ref.~\onlinecite{ni2mnsn} using the
MFA and the FM reference state is 323~K, if 
only the Mn-Mn exchange integrals are considered.
Our related test value is 315~K.
The Curie temperature rises to 360~K if  Ni-Mn exchange
interactions are also included \cite{ni2mnsn}.
Our mean-field value based on the DLM reference state is 373~K.
We would like to point out that good agreement of the Curie temperature
obtained in Ref.~\onlinecite{ni2mnsn} from the FM reference 
state with the experiment is probably fortuitous, as the RPA would lower 
this value. The present RPA value based on the DLM reference 
state (322~K) agrees well with the experiment.

\subsubsection{Extraction of exchange interactions from experiment}
\begin{figure*}
\center \includegraphics[width=18cm]{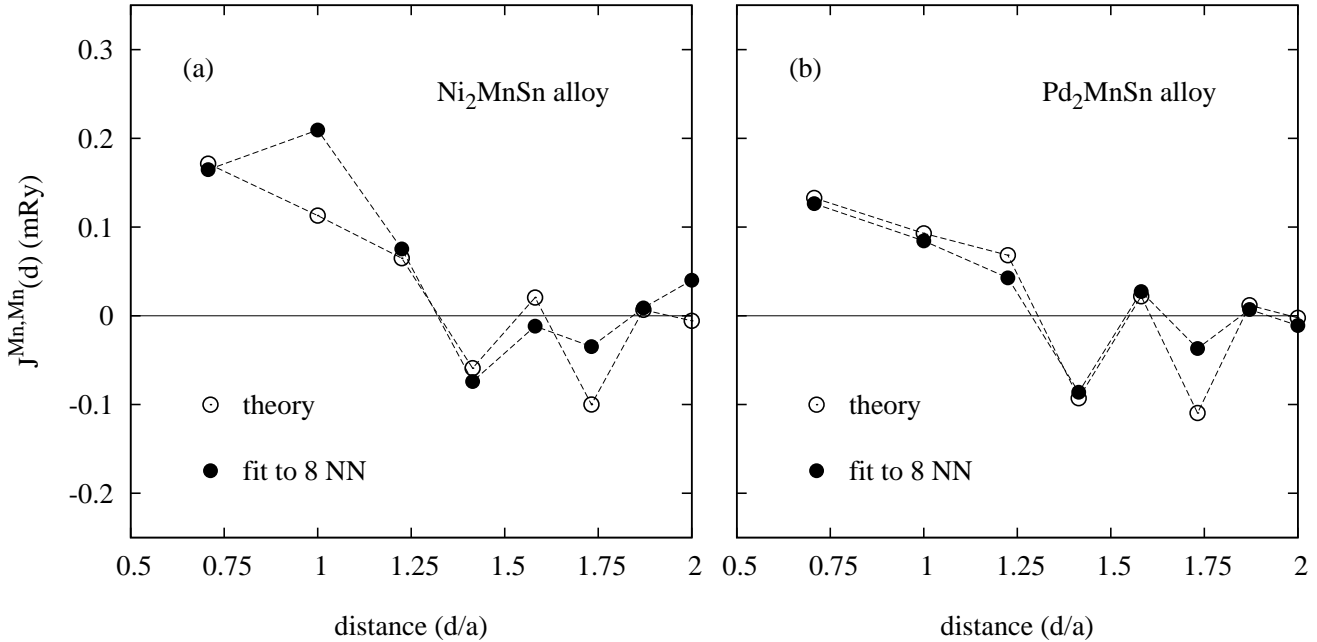}
\caption {Comparison of calculated exchange integrals
(present paper, theory) with those obtained from the
fit to neutron spin-wave scattering experiments
(Table~I of Ref.~\onlinecite{expfit}, fit to 8 NN):
(a) Ni$_{2}$MnSn, and (b) Pd$_{2}$MnSn.
}
\label{fJe}
\end{figure*}

 Noda and Ishikawa \cite{expfit} have extracted up
to 8 nearest-neighbor (NN) exchange interactions for Ni$_{2}$MnSn and
Pd$_{2}$MnSn from the fit to measured inelastic neutron spin-wave scattering data.
Taking care of the fact that the authors' exchange integrals do not
include spin moments and different definitions of the prefactor in
the Heisenberg Hamiltonian, we compare in Fig.~\ref{fJe} our calculated results with the
values extracted by Noda and Ishikawa \cite{expfit}.
The general agreement between the calculated results and the extracted values from the
fit appears to be good.
It should be noted that calculated exchange integrals are not
limited by the distance 2~$a$ (8 NN) as the fitted ones.
In fact, saturation of results with respect to the number of shells  is not achieved
 for 8 NN interactions (see Fig.~\ref{f5}).
It is also clear from  Table~I of Ref.~\onlinecite{expfit}
that the fitted values can change significantly depending on the
number of exchange integrals used for a fit.
For example, in case of  Ni$_{2}$MnSn the 2nd neighbor interaction is larger than that of 
the 1st neighbor for the 8 NN fit, and smaller for the 6 NN fit.

Ideally, one expects the fitted values to decrease in their
sizes.  This  is not so for a relatively large $J(8)$ in Table~I
for Ni$_{2}$MnSn \cite{expfit} which, for an optimal fit, should
be of  the smallest magnitude.
This indicates that the fitted values should not be taken too
literally.  Seemingly perfect agreement between the measured magnon spectrum and 
that calculated from the fitted values for Pd$_{2}$MnSn in Ref.\onlinecite{expfit}
does not guarantee similar agreement in other related magnetic properties.

We  also estimate the spin-stiffness constants $D_{\rm stiff}$
for Ni$_{2}$MnSn and Pd$_{2}$MnSn.
The spin-stiffness is a property of the ferromgnetic state,
i.e., the state at T=0~K.
We have therefore estimated $D_{\rm stiff}$ from calculated
Mn-Mn exchange integrals in the FM state.
The real-space expression for $D_{\rm stiff}$ is  non-convergent,
and a special treatment is needed for its determination \cite{pajda}.
 Estimated values of $D_{\rm stiff}$ for Ni$_{2}$MnSn and
Pd$_{2}$MnSn are 160~$\pm 25$ (150~$\pm 10$)~meV.\AA$^{2}$ and
130~$\pm 25$ (98~$\pm 10$)~meV.\AA$^{2}$, respectively.
The values in brackets are experimental values \cite{expfit}, and
the agreement between the theory and experiment is acceptable.
The theoretical error bars were estimated from differences in
calculated $D_{\rm stiff}$ for various sets of damping constants
\cite{pajda}.

\begin{figure}
\center \includegraphics[width=8.5cm]{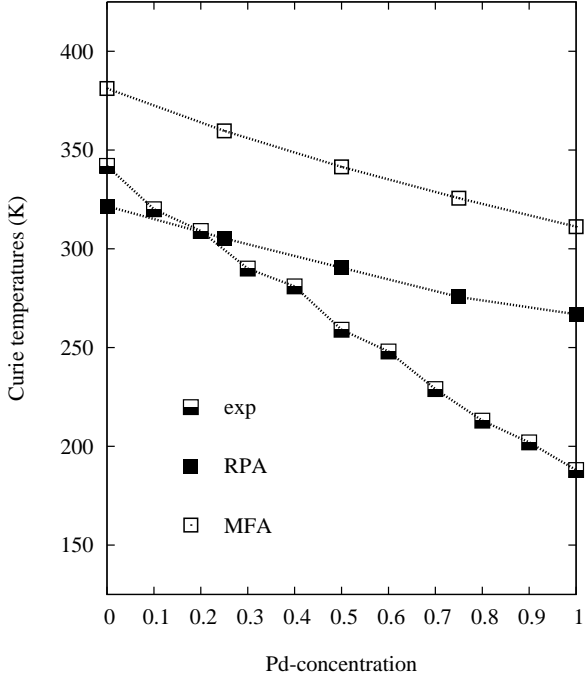}
\caption {The same as in Fig.~\ref{f7} but for
(Ni$_{1-x}$,Pd$_{x}$)$_{2}$MnSn. The calculated values (RPA)
are compared with experimental data \cite{qha2}. The MFA
values of Curie temperatures are also shown.
}
\label{f8}
\end{figure}

Finally, corresponding results for (Ni$_{1-x}$,Pd$_{x}$)$_{2}$MnSn
alloy are shown in Fig.~\ref{f8}.
The experimental values\cite{qha2} of the Curie temperature lie in the
range 190-340~K.
In line with the concentration dependence of the exchange integrals
(see Fig.~\ref{f5}) one expects a linear decrease of the Curie
temperature with the Pd-content.
The calculated RPA results  show an almost linear decrease
of the Curie temperature found in the experiment, although the
decrease is smaller than what is observed experimentally\cite{qha2}.
The MFA values are systematically above the RPA as well as the
experimental results, but do show the linear decrease of the Curie
temperature.
Some disorder in Heusler alloys, even for stoichiometric compound, is
rather frequent.
In the case of Pd$_{2}$MnSn alloy one can speculate about some disorder
between Mn- and Sn-sites, which could reduce the calculated
Curie temperature (due to antiparallel alignment of Mn-moments). 
Such a study would require a generalization of the RPA to
random systems, which is beyond the scope of the present article.

\subsection{Resistivity}
\label{RR}

In general, there is very little known about the resistivity of Heusler alloys, 
either experimentally or theoretically.
 In magnetic alloys there are three contributions
to the resistivity: 
a temperature-independent part,  residual 
resistivity due to the alloy disorder and other defects, and two temperature-dependent 
terms due to electron-phonon  and electron-magnon scatterings.

\subsubsection{Residual resistivity}

\begin{figure}
\center \includegraphics[width=8.5cm]{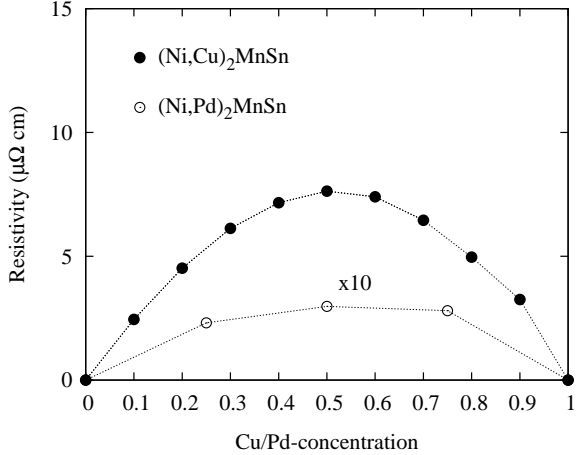}
\caption {The concentration dependence of residual resistivities 
in (Ni$_{1-x}$,T$_{x}$)$_{2}$MnSn alloys (T=Cu, Pd) evaluated
for the ferromagnetic ground state at the zero temperature.
Note the factor 10 multiplying resistivities for T=Pd alloy.
}
\label{f9}
\end{figure}

The residual resistivities are calculated using FM reference state, appropriate for zero temperature.
 The calculated resistivities for (Ni$_{1-x}$,T$_{x}$)$_{2}$MnSn,
T=Cu, Pd are shown in Fig.\ref{f9}.
The result, an almost parabolic concentration dependence of residual
resistivities (Nordheim rule \cite{resT}) for both alloys, is 
in agreement with our discussion of the character of the disorder 
at the Fermi energy (see Sec.~\ref{DOS}), namely its weakness.
Much smaller resistivity of (Ni,Pd)$_{2}$MnSn alloy is due to 
the site off-diagonal disorder,
whose effect on the resistivity is  weaker than that of the diagonal disorder which 
dominates in case of (Ni,Cu)$_{2}$MnSn.

\subsubsection{Temperature-dependent resistivity} 

The contribution due to phonons is relatively well understood \cite{resT}, and it is
small in Heusler alloys for temperatures around the room temperature 
\cite{resT1,resT2}.
The most important contribution to the temperature-dependent resistivity
is due to magnon scattering or the spin-disorder resistivity. 
A simple theoretical model of the spin-disorder induced resistivity
of magnetic alloys was developed some time ago by Kasuya \cite{spindis}.
This theory explains the experimental facts that for temperatures higher 
than $T_c$ the resistivity is almost constant,
while for temperatures below $T_c$ the resistivity
decreases and it is equal to the residual resistivity at zero temperature.
The increase of the resistivity with temperature (above $T=0$ and below $T_c$) is due to the 
increasing amount of the spin disorder.
The spin-disorder itself is best characterized by the spin-spin 
correlation function \cite{spindis2}, which can be determined from 
first-principles using known exchange integrals
(see e.g. Ref.~\onlinecite{spinspin}).
Here we  adopt a less ambitious approach. 
For temperatures above $T_c$, we assume that 
 a reasonable model of spin-disorder is simply the DLM
model, which assumes that the local
magnetic moments on  atoms are oriented randomly with equal probabilities 
in all directions. The net moment is zero and  the spins are uncorrelated 
(the spin-spin correlation function is zero in this case).  
The DLM model is  formally equivalent to a random alloy problem \cite{dlm}, and the 
TB-LMTO-CPA code employing the Kubo Greenwood
formula can be used conveniently to compute the resistivity.

\begin{figure}
\center \includegraphics[width=8.5cm]{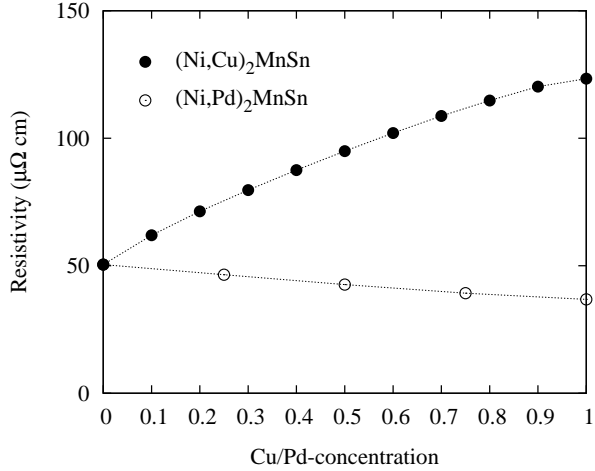}
\caption {The concentration dependence of the resistivity of
(Ni$_{1-x}$,T$_{x}$)$_{2}$MnSn alloys (T=Cu, Pd) due to
the spin-disorder evaluated for temperatures above the Curie
temperature. Magnetic state is represented by the disordered
     local moment (DLM) state.
}
\label{f10}
\end{figure}

Schreiner \etal \cite{resT1} studied the temperature dependence of resistivity of
 a series of Heusler alloys including
Ni$_{2}$MnSn and Pd$_{2}$MnSn \cite{resT1}.
The resistivities calculated for the DLM reference states 
for (Ni,T)$_{2}$MnSn (T=Cu,Pd) alloys 
are presented in Fig.~\ref{f10}.
We remind the reader that this resistivity accounts only for the 
spin-disorder and does not include the part due to scattering from phonons.
We observe an increase of the resistivity due to spin-scattering for
Cu-impurities in the host Ni$_{2}$MnSn and a decrease for Pd impurities.
Fig.~\ref{f10} reflects  the decreasing degree of localization
(correspondingly, increasing Mn-X hybridization) along the sequence
Cu-Ni-Pd, which leads to a decrease of the relative strength of the
magnetic disorder, and therefore resistivity, along the same sequence.
The corresponding total resistivities measured in the experiment for 
the ordered Ni$_{2}$MnSn and Pd$_{2}$MnSn alloys are about 
75~ $\mu$Ohm.cm and 50~$\mu$Ohm.cm, respectively.
The experimental values obtained after subtraction of the contribution
from phonon scatterings are 47~$\mu$Ohm.cm and 22$-$30~$\mu$Ohm.cm.
These values compare reasonably well with the values 50~$\mu$Ohm.cm and 
37~$\mu$Ohm.cm obtained from the present simple theory.
In particular, larger value of the spin-disorder induced part of the
resistivity in Ni$_{2}$MnSn as compared to Pd$_{2}$MnSn is correctly
reproduced.

We present a simplified treatment of the temperature dependence of 
resistivity due to spin disorder for temperatures below $T_c$ \cite{spindis,spindis2}.
 We use two simple approaches, based on  (i)
simulation of the temperature dependence through the uncompensated DLM
(uDLM) model \cite{akai,magall}, and (ii) assuming experimentally
observed quadratic temperature dependence of the resistivity due
to spin-disorder in combination with calculated values of $T_c$
 and the resistivity above the Curie temperature.

\begin{figure}
\center \includegraphics[angle=270,width=10cm]{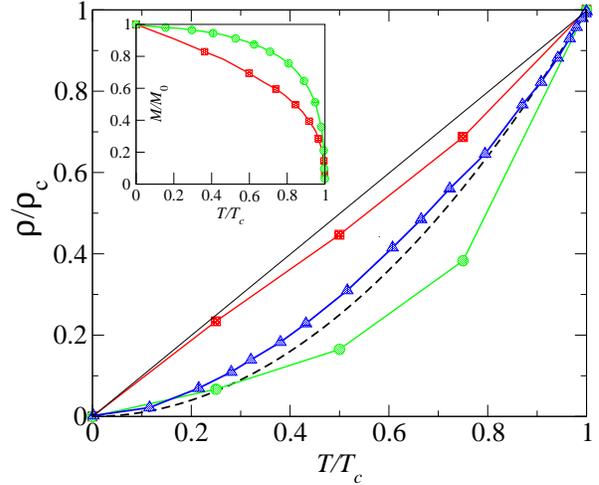}
\caption {(Color online) Reduced resistivity $\rho({ T})/\rho_{c}$ vs reduced
temperature $T/T_{c}$ (the suffix $c$ relates to the Curie
temperature) for two models of the temperature dependence
of the magnetization: (i) as the temperature dependence
of the Sn hyperfine field in Ni$_{2}$MnSn (squares, Ref.\onlinecite{hf-sn}), and
(ii) as the temperature dependence of the Fe hyperfine
field in bcc-Fe (circles, Ref.\onlinecite{hf-fe}). The experimental values including
the phonon part (triangles) are from. Ref.~\onlinecite{resT1}. The dashed line is a simple quadratic curve,
showing the magnon-part of the reduced resistivity. Being in the reduced form, it is actually free of the
constant $B$: $\rho/\rho_c = (T/T_c)^2$.
 The straight line is used as
a guide for eyes. The inset shows the variation of the reduced magnetization $M/M_0$ as a function of $T/T_c$
 determined from Ref.\onlinecite{hf-sn} (squares) and Ref.\onlinecite{hf-fe}
(circles). $M_0$ is the saturation magnetization.}
\label{fR1}
\end{figure}

The results of the first model are shown in Fig.~\ref{fR1}.
For a given temperature below $T_c$ we identify the
magnetization of the system with that corresponding to the uDLM model
with a specific ratio of the parallel and antiparallel spins.  We calculate
the resistivity for this ratio using the conventional Kubo-Greenwood approach.
If there are no antiparallel spins in the the system we
have the FM state corresponding to the temperature T=0~K.
On the other hand, for an equal number of parallel and antiparallel
spins we have the DLM state with zero total magnetization, corresponding
to resistivity at and above the Curie temperature.

Optimally one can determine the magnetization of the system from the
present Heisenberg Hamiltonian using statistical mechanical methods.
Alternatively, one can adopt some models.  
We used two, namely the
temperature dependence of the magnetization as obtained from the
hyperfine field (HF) measurements on Sn atoms in Ni$_{2}$MnSn \cite{hf-sn}
and that obtained from similar measurements on Fe-atoms in bcc-Fe
\cite{hf-fe}. 
From the reduced magnetization $M/M_0$ at a given reduced temperature $T/T_c$ we determine the 
ratio of parallel and antiparallel spins for the uDLM model, and associate the 
corresponding reduced resistivity $\rho/\rho_c$ with the reduced temperature $T/T_c$. 
For a typical ferromagnetic$\leftrightarrow$paramagnetic transition the reduced magnetization curve  close
to $T/T_c$=1 has a universal (system-independent) form. So our results for $\rho/\rho_c$ obtained
this way should be more reliable in this range. A couple of comments are in order
at this point. 
The HF field at Sn atoms in Ni$_{2}$MnSn
is not directly proportional to the magnetization. 
Several models for the dependence of the HF field
on the magnetization are discussed in the literature 
and one such model was used in Ref.~\onlinecite{hf-sn}
by Gavriliuk \etal 
The seemingly linear decline of magnetization with temperature around $T=0$ contradicts
the Bloch $T^{3/2}$ law, which a typical ferromagnetic material is expected to obey.
It is in view of this and 
the uncertainties involved in relating the HF field at Sn atoms to
the magnetization that we include the
second model based on bcc Fe as the prototype of a ferromagnetic$\leftrightarrow$paramagnetic transition.
The actual magnetization vs. temperature curve for Ni$_2$MnSn is expected to lie between the red (squares)
and the green (circles) curves shown in the inset of Fig.~\ref{fR1}.
 A typical convex form of the temperature dependence of the reduced resistivity
is well reproduced (see Fig.~\ref{fR1}) for both cases.
The experimental curve including the phonon contribution lies in between these two simple model curves.
For comparison, in Fig.~\ref{fR1} we also include the quadratic form $\rho/\rho_c = (T/T_c)^2$ (dashed curve), based on the
experimental observation that the magnon resistivity has a simple quadratic dependence on temperature \cite{resT1}.
The difference between the quadratic form (dashed curve) and the experimental curve is the contribution due
to the phonons and we see that it is quite small.

\begin{figure}
\center \includegraphics[angle=270,width=10cm]{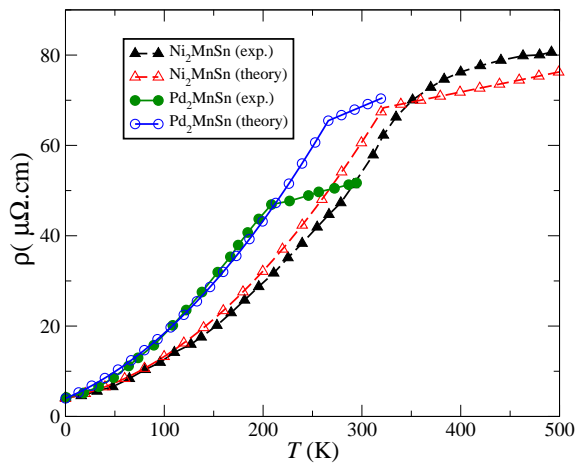}
\caption {(Color online) The temperature dependence of the resistivity
determined as described in the text for Ni$_{2}$MnSn
(triangles) and Pd$_{2}$MnSn (circles). The empty and full
symbols denote  the calculated and experimental \cite{resT1} results,
respectively.
}
\label{fR2}
\end{figure}

 In Fig.~\ref{fR2} we display theoretically calculated
resistivity assuming the  form \cite{resT1}
${\rho({ T})}$= ~${\rho_{0}}~+~{ A\,T~+~B\,T^{2}}$.
The quadratic term is due to the spin disorder \cite{spindis,spindis2},
the linear one is the phonon contribution (which is valid with exception
of very low temperatures \cite{resT}), and $\rho_{0}$ is the value of
the residual resistivity at $T$=0~K, which is due to all the defects
present in the sample.
While the values of ${\rho_{0}}$ and the coefficient $A$ were taken from
the experiment \cite{resT1}, the coefficient $B$ was determined from the
calculated $T_c$ and the resistivity in the DLM state.
We obtain  good agreement between the theory and experiment, in particular
for Ni$_{2}$MnSn.
It should be noted that some offsets of the  calculated curves as compared
to the experiment are due to differences in calculated and measured
$T_c$ which is larger for Pd$_{2}$MnSn.
Theoretical estimates of the coefficient $B$ are 4.79- and 
5.2$\times$10$^{-4} \mu\Omega$.cm K$^{-2}$
for Ni$_2$MnSn and Pd$_2$MnSn, respectively. 
The corresponding experimental values are 3.94- 
and 6.1$\times$10$^{-4} \mu\Omega$.cm K$^{-2}$.
Considering the fact that $B$ represents the second 
derivative of resistivity with respect to temperature,
this agreement is good, as evidenced by Fig.\ref{fR2}.
\section{Conclusions}
\label{Con}

We have studied the electronic, magnetic, thermodynamic, 
and transport properties of quaternary Heusler alloys (Ni,T)$_{2}$MnSn
(T=Cu, Pd) by means of  first-principles  
density functional method.
In agreement with experiments, magnetic moments per formula unit 
 depend only weakly on the alloy composition for both alloys,  
having values around 4~$\mu_{\rm B}$.
Exchange interactions were determined using the DLM reference state,
 which assumes no \textit{a priori} magnetic ordering in the system.
The Curie temperatures were estimated using RPA applied to the non-random Mn-sublattice.
The alloy disorder strongly influences values of exchange integrals 
in (Ni,Cu)$_{2}$MnSn alloys, while only weak concentration dependence
of exchange integrals is found for (Ni,Pd)$_{2}$MnSn alloys.
Consequently, a linear decrease of the calculated Curie temperature
is obtained in the latter alloys in agreement with the experiment.
A more complex concentration dependence of exchange integrals in 
(Ni,Cu)$_{2}$MnSn alloys, in particular different behaviors of Ni-rich
and Cu-rich alloys, can be ascribed to the 3rd NN exchange integrals.
This result also confirms a model of Stearns \cite{stearns1} on the 
relevance of the first three exchange integrals for magnetic properties
of Heusler alloys.
In fact, a much larger number of exchange integrals is needed to obtain
reasonably stable results in \textit{ab initio} theoretical studies.
This is in striking contrast with the related semi-Heusler alloy (Cu,Ni)MnSb,\cite{qsha1}
where the exchange interactions are strongly damped due to their halfmetallicity.

The residual resistivities obey the weak-scattering Nordheim rule. This is due to
the fact that strong disorder found e.g. in (Ni,Cu)$_{2}$MnSn alloys
influences energy states far from the Fermi energy relevant for electronic transport.
Using a simple model for the spin-disorder, we have  estimated 
the temperature dependent resistivity at temperatures above $T_c$. 
 Reasonably good agreement with  experimental results is found for calculations which assume that
 the resistivity above $T_c$ is essentially captured by the DLM model.
Using this value of resistivity and calculated Curie temperatures, good agreement
with the experiment was obtained also for the temperature dependence
of the resistivity below the Curie temperature.

\begin{acknowledgments}
This work was supported by a grant from the Natural Sciences and Engineering Research Council of Canada.
J.K. and V.D. acknowledge financial support from AV0Z 10100520 and the  Czech Science Foundation (202/09/0775).
The work of I.T. has been supported by the Ministry of Education
of the Czech Republic (Project No. MSM 0021620834) and by the Czech Science Foundation (Project No. 202/09/0030).
J.K. also acknowledges the kind hospitality of the Physics Department at Brock University where most of this work was carried out.

\end{acknowledgments}

\end{document}